\documentclass[prl,showpacs,twocolumn]{revtex4}
\usepackage{graphicx}   
\usepackage{dcolumn}    
\usepackage{bm}         

\begin{document}
\title{Bragg Spectroscopy of Vortex Lattices in Bose-Einstein
condensates}
\author{S. R. Muniz}
\email{sergio.muniz@physics.gatech.edu}
\author{D. S. Naik}
\author{C. Raman}
\affiliation{School of Physics, Georgia Institute of Technology, Atlanta, Georgia 30332}
\date{\today}

\begin{abstract}
We have measured the velocity field of a vortex lattice within a sodium Bose-Einstein condensate using Bragg scattering. The phase gradient of the macroscopic wavefunction was mapped into the spatial structure of the diffracted atom cloud, allowing for single shot measurement of the rotation parameters. A combination of spectral and spatial information yields a complete description of the superfluid flow, coarse-grained over the lattice structure, including direct and independent measurements of the rate and sense of rotation. Signatures of the microscopic quantum rotation have also been observed
\end{abstract}

\maketitle

\begin{figure}[tbph]
\begin{center}
\vspace{-3mm}
\includegraphics[width = 0.37 \textwidth]{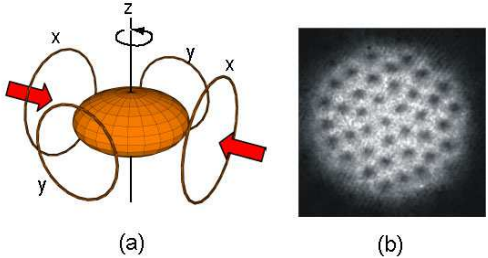}
\end{center}
\vspace{-8mm}
\caption{(Color online) Bragg scattering from vortex lattices.  (a) Atoms in a TOP trap are made to rotate about the $z$-axis by phase control of the transverse fields produced in coil pairs $x$ and $y$. This procedure forms a lattice of quantized vortices (b) which is imaged in time-of-flight after 40 ms.  The field of view is 0.6 mm.  The Bragg beam containing frequencies $\omega$ and $\omega+\delta$ is applied along the $x$-direction and retroreflected, as shown in (a).}
\vspace{-2mm}
\label{Figure1: Vortex-Lattice}
\end{figure}

Vortices are a cornerstone in the study of superfluids and have applications throughout fluid mechanics and condensed matter physics \cite{donn91,bare01,nels01}.  Recently, gaseous Bose-Einstein condensates under rotation have become an important test bed for predictions of the behavior of quantized vortices, which form highly regular lattices \cite{madi00,abos01latt,halj01,hodb01}. In these gases, time-of-flight (TOF) imaging has proved instrumental for observing the rotation of the cloud through the detection of the vortex cores (see Figure \ref{Figure1: Vortex-Lattice}b for an example). However, this technique only measures the superfluid density, and not the phase of the macroscopic wavefunction. Phase measurements would provide a more complete picture of vortex states, especially useful in cases where individual vortices cannot be readily detected or where the TOF expansion dynamics might be unknown at first. In atomic gases, powerful optical and spectroscopic techniques \cite{kozu99bragg,sten99brag,sims00,thei04,katz04} can be readily applied, which afford new possibilities for performing phase measurements.

In this work, we have used two-photon Bragg scattering to study vortex lattices in a BEC.  This technique directly probes the velocity distribution of the rotating superfluid, which is the gradient of the wavefunction phase. Bragg scattering has been proposed for detecting the phase singularity of a single vortex \cite{blak01} and was used to measure the spatial phase of a dark soliton \cite{bong03}. Earlier work \cite{sten99brag,rich03,thei04} has explored the spectroscopic nature of this tool for condensates; however, that method requires repeated experimental runs to obtain an entire spectrum. In this letter we demonstrate that Bragg scattering provides a unique window into a rapidly rotating BEC through the {\em spatial profile} of the diffracted atoms.  A significant amount of information can be gleaned from a single image \cite{ozer02}. In particular, one can directly extract the rate and the sense of rotation from the TOF image.We use spectral and spatial information to extract a complete picture of the two-dimensional velocity flow.


In our experiments we produce a sodium BEC with typically $5-7\times 10^{6}$ atoms in a optically plugged quadrupole magnetic trap (OPT) following the procedures described in reference \cite{naik05}. After reaching quantum degeneracy the atoms are transferred into a time-averaged orbiting potential (TOP) trap \cite{petr95}, which has the zero-crossing of the magnetic field rotating in the horizontal plane. The transfer procedure occurs by linearly ramping down the intensity of the blue detuned optical plug within 200 ms, after which we suddenly turn on a rotating bias field. Following the transfer, the atoms are allowed to equilibrate via saddle-point evaporation in the TOP trap for a hold time of 1 sec.  At the end of the process, we typically have $1-3\times 10^{6}$ atoms in an almost pure condensate (typically larger than $90\%$). We used a radial gradient  $B_\rho' = 12 $ Gauss/cm and a bias rotation of $\omega _{TOP}=2\pi\times5$ kHz. The measured transverse oscillation frequency is $\omega_{\rho}=2\pi \times 31$ Hz, with $\omega_z =\sqrt{8} \omega_{\rho}$.

We produce the vortex-lattice by temporarily creating a rotating elliptical asymmetry in the horizontal \textit{x-y} plane of the TOP trap \cite{hodb01}. This is done by combining the signals of two digital frequency synthesizers operating at frequencies $\omega_{1} = \omega_{TOP}+ \omega_{AR}$ and $\omega_{2} = \omega_{TOP}- \omega_{AR}$.  These signals are each split, phase shifted and summed together to produce a slow rotation superimposed on a fast one:  $B_x(t) = B_0 \cos(\omega_1 t)+\epsilon \cos(\omega_2 t)$ and $B_y(t) = B_0 \sin(\omega_1 t)-\epsilon \sin(\omega_2 t)$, where $\epsilon$ and $\omega_{AR}$ are the amplitude and frequency of the rotating asymmetry, respectively.
The two currents are individually amplified using 100 Watt car audio amplifiers, and capacitively coupled to a pair of Helmholtz coils of approximately 10 cm diameter along the $x$ and $y$ directions, respectively.  In order to maximize the number of vortices, we chose $\omega_{AR}= 2\pi \times 22$ Hz, which is very close to the frequency $\simeq 0.7 \omega$ that drives the quadrupole mode in our harmonic trap \cite{madi01}. After applying the rotating asymmetry for 1.5 seconds, it was turned off and the atomic cloud allowed to equilibrate in the radially symmetric TOP trap for 1 to 1.5 seconds. This procedure reliably created lattices with approximately 40$\pm$10 vortices, as shown in Fig. \ref{Figure1: Vortex-Lattice}b.

After producing vortices and allowing the lattice to equilibrate, we pulse the Bragg diffracting beams along the $x$-direction (see Figure \ref{Figure1: Vortex-Lattice}a) for a time $T_{B}$, while the atoms are still in the trap. The Bragg beams are detuned by 1.7 GHz from the $F=1$ to $F'=2$ resonance, and are created by back-reflecting a single beam that contains two frequencies $\omega_{L}$ and $\omega_{L} +  2\pi \delta$.  $\delta$ is the difference between the frequencies of two rf synthesizers that are used to drive a single acousto-optic modulator.  The interaction of the matter wave with the moving optical lattice produces two diffracted (outcoupled) clouds of atoms, one with momentum $+\textbf{q}$ and the other with $-\textbf{q}$.  In the free particle regime, the Bragg resonance condition is given by $\delta^{fp}_{o} = \frac{q^2}{2 m h} + \frac{\textbf{q} \cdot \textbf{v}}{h}$ and expresses the conservation of momentum and energy of the two-photon process. The second term in $\delta^{fp}_{o}$ is simply the Doppler shift, which makes the Bragg technique velocity sensitive. For sodium atoms near the principal resonance, $\frac{q^2}{2 m h}=100$ kHz.  For a trapped BEC, one also has to consider the effects of interactions. In the mean-field and local density approximations \cite{zamb00, blak02}, this causes an extra $\delta_{MF} = \frac{4 \mu}{7 h}$ frequency shift \cite{sten99brag}.  In our case, for a stationary condensate, the Bragg resonance is peaked at a frequency $\delta_o = \frac{q^2}{2 m h} + \delta_{MF} \approx 101$ kHz.  Here, $q=|\textbf{q}| = 2h / \lambda$ is the transferred momentum, $\mu$ is the chemical potential, $m$ and $\textbf{v}$ are the atomic mass and velocity.

\begin{figure}[tbph]
\begin{center}
\includegraphics[width = 0.43 \textwidth]{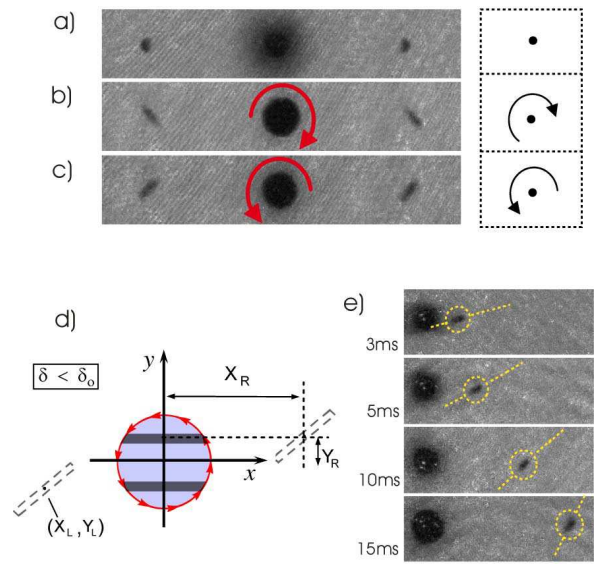}
\end{center}
\caption{(Color online) Vortices probed by Bragg scattering.  Outcoupled atoms (to the far right and left of each image) showed no particular structure for non-rotating clouds (a), whereas from vortices (b) and (c), the outcoupled atoms were tilted according to the direction of rotation. Also in (a), the higher atomic density causes collisional induced heating and makes the cloud look more diffuse. Images (a-c) were taken at 10 ms TOF.  The tilt angle increases with respect to time of flight, as shown in (e). Each pair of Bragg frequencies is resonant with a thin strip of atoms parallel to the $x$-axis, as illustrated in (d). All images were taken at $\delta = 102$kHz.}
\vspace{-5mm}
\label{Figure2: Spatial-Structures}
\end{figure}

We applied a square-shape Bragg pulse with $T_{B}=250 \mu s$ and then turned off the magnetic trap within $100 \mu$s. The atoms were allowed to expand for a variable time-of-flight $t_{tof}$ before taking an absorption image using laser light resonant with the $F=1\rightarrow2$ transition in a $250\mu$s pulse.  Our main result is shown in Figure \ref{Figure2: Spatial-Structures}, where we can clearly observe {\em spatial structures} in the outcoupled atom cloud arising from the rotation of the lattice.  Figure \ref{Figure2: Spatial-Structures}a shows the diffraction from an initially stationary condensate, and the outcoupled atoms appear to the right and left of the stationary condensate.  No particular structure is visible.  However, in Figure \ref{Figure2: Spatial-Structures}b, we have initially prepared a vortex lattice, which causes the diffracted atoms to form a tilted, elongated spatial pattern.  Moreover, when we reversed the direction of the applied rotation (by replacing $\omega_{AR}\rightarrow -\omega_{AR}$), the tilt angle with respect to the $y$-direction reverses, as shown in Figure \ref{Figure2: Spatial-Structures}c.

We can understand our observations in terms of a simple picture of rigid body rotation, which applies for condensates containing a large number of vortices. The classical behavior results from a coarse-grained average over the velocity field of several vortices. Even though locally the tangential velocity around each core axis varies $\propto 1/r$, the bulk fluid velocity follows $v = \Omega r$, with $\Omega = |\vec{\Omega}|$ proportional to the number of vortices.

The Bragg process selects a group of atoms with the same projection of velocity $v_x$ along the direction of the momentum transfer. In the rigid-body limit, since $v_x = \Omega y$, the resonance condition is given by $\delta = \delta_o + 2y\Omega / \lambda$. Therefore, for a spectrally narrow Bragg pulse, with $\delta <\delta_0$ and a counter-clockwise rotation, the resonance corresponds to a thin, horizontal band of atoms with $y>0$ for atoms which are Bragg scattered to the right, and $y<0$ for atoms scattered to the left (the dark shaded regions in Figure \ref{Figure2: Spatial-Structures}d). In ballistic expansion, $v_y$ results in a tilted stripe whose angle increases with time, as observed in Figure \ref{Figure2: Spatial-Structures}e. At long TOF the stripe should become fully stretched along the vertical axis of the images.

\begin{figure}[tbph]
\begin{center}
\includegraphics[width = 0.4 \textwidth]{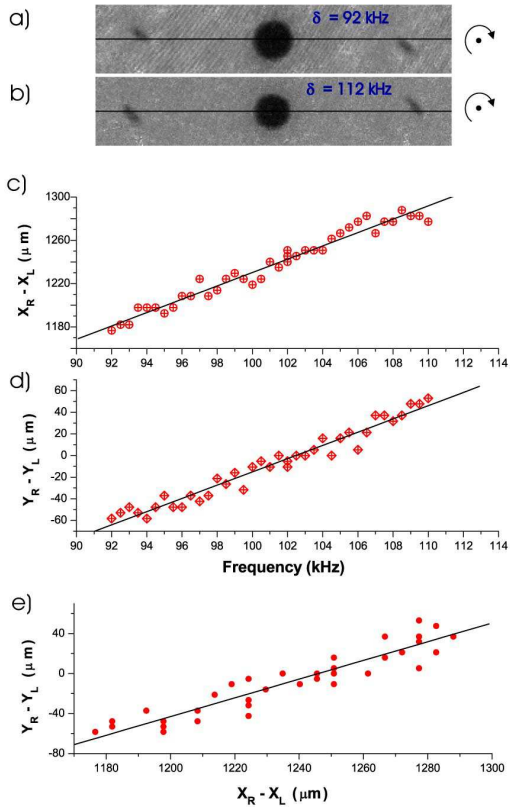}
\end{center}
\vspace{-9mm}
\caption{(Color online) Mapping the velocity field of a vortex lattice.  Measuring the center-of-mass locations of the right and left diffracted atoms ( $(X_R, Y_R)$ and $(X_L,Y_L)$, respectively, as defined in Figure \ref{Figure2: Spatial-Structures}) in the $x-y$ plane allows us to reconstruct the velocity profile of the rotating cloud.  For negative detuning, as in (a),  $\delta-\delta_0 =-10$ kHz, and therefore $Y_L>0$ and $Y_R<0$.  For positive detuning, the opposite is the case, as in (b) where $\delta-\delta_0 = +10$ kHz.  The difference $X_R-X_L$ is mapped as a function of frequency in (c), while $Y_R-Y_L$ is plotted in (d).  Finally, in (e), we have plotted $Y_R-Y_L$ versus $X_R-X_L$ for the data in (c) and (d).  The data in (e) is a direct map of the position ($Y$-coordinate) versus velocity ($X$-coordinate) distribution within the vortex lattice, and shows a linear relationship consistent with rigid body rotation.  The lines are fits to the data}
\vspace{-2mm}
\label{Figure3: CM Displacements}
\end{figure}

According to this model, the vertical position of the center of mass for the right (left) diffracted cloud $Y_{R}$  ($Y_{L}$) is set by the Bragg resonance condition, and should depend on the detuning $\delta$. Likewise, the horizontal position $X_R$ ($X_L$) is given by the $(\textbf{q} \cdot \hat{\textbf{x}}/m + v_x)t_{tof}$ and, due to its dependence on $v_x$, it also depends on $\delta$ . As a result, for a clockwise rotation and $(\delta - \delta_o) < 0$, we should have $Y_R<0$ and $Y_L>0$. These predictions are clearly verified in Figure \ref{Figure3: CM Displacements}a, for $t_{tof} = 10$ ms, at $(\delta - \delta_o) = -10$ kHz. For $\delta = $+10 kHz, the opposite is true, namely, $Y_R>0$ and $Y_L<0$, as seen in Figure 3b. The coordinate pair $(X,Y)$ is the center of mass of the diffracted atom cloud within each image, as determined from a parabolic (Thomas-Fermi) fit to the data.

It is straightforward to show that the relative displacements between each scattered group are given by $\Delta$Y$=(Y_R-Y_L)=\frac{\lambda}{\Omega} (\delta - \delta_o)$ and $\Delta$X$=(X_R-X_L)=2\times(\frac{2 h}{\lambda m} + \lambda \delta )t_{tof}$. Figures \ref{Figure3: CM Displacements}c to \ref{Figure3: CM Displacements}e show that our measurements are in very good agreement with the expected linear behavior for this model. In particular, while \ref{Figure3: CM Displacements}c is simply a measure of the Doppler sensitivity of the technique, Figures 3d and 3e clearly manifest the rigid body rotation of the cloud.  Furthermore, Figure 3e provides the complete mapping of the velocity profile of the condensate.  The straight-line dependence of $\Delta Y$ vs. $\Delta X$ is a direct consequence of the velocity field $\vec{v} = \vec{\Omega} \times \vec{r}$.

A least square linear fitting of the data in Figure \ref{Figure3: CM Displacements}e leads to a macroscopic angular frequency $\Omega=2\pi \times( 14.9 \pm 0.8$ Hz). We also calculated the rotation frequency from the evolution of the tilt angle as function of time (see Fig. \ref{Figure2: Spatial-Structures}e), $\theta(t)=\arctan( \Omega t)$  , resulting in $\Omega=2\pi \times( 15.4 \pm 1.1$ Hz). In addition, we used a third independent method to estimate $\Omega$, based on the total quantized vorticity of the lattice. For that we noted the fact that in the rigid body limit $\Omega= (h N_V)/(2 m \pi R_{\rho}^2)$, and therefore, by measuring the number of vortices $N_V$ one can calculate the rotation rate. In our case $R_{\rho}=35\mu$m is the calculated Thomas-Fermi radius in the \emph{x-y} plane, and $N_V$ was determined by a manual counting of the number of vortices from several images taken at long TOF. This resulted in $N_V = 37 \pm 7$, which leads to $\Omega=2\pi \times( 13.3 \pm 2.6$ Hz$) $. An apparent undercounting in the number of vortices has also been previously noted in ref.\ \cite{Halj03thesis} and appears to be consistent with some numerical and analytical calculations \cite{fede01prl2,angl02vortex}.
All the methods produced similar results for $\Omega$ and indicate that the condensate was not completely equilibrated with the rotating potential at frequency $\omega_{AR}$. However, we observed that longer rotation times only decreased the number of vortices, probably due to some residual asymmetry in the trapping frequencies of our TOP trap.

\begin{figure}[tbph]
\begin{center}
\includegraphics[width = 0.42 \textwidth]{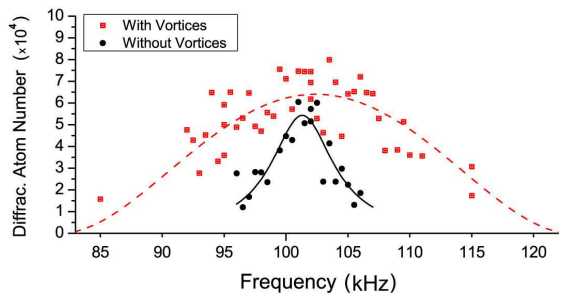}
\end{center}
\vspace{-7mm}
\caption{(Color online) Spectroscopy of the vortex lattice. The number of Bragg diffracted atoms is measured as function of the frequency \textbf{$\delta$}, for samples with (\emph{solid circles}) and without (\emph{squares}) vortices. A significant broadening of the spectrum is observed in the presence of vortices due to the broad velocity distribution caused by the rigid-body rotation. The solid curve represent the best Lorentzian fit to the data, while the dashed curve refers to the model described in the text.}
\vspace{-2mm}
\label{Figure4: Bragg-spectrum}
\end{figure}

In addition to this spatial analysis, we can also use Bragg spectroscopy to derive information about the velocity profile from the {\em number} of outcoupled atoms \cite{sten99brag,bong03}. We measured this number as a function of the frequency $\delta$, with and without vortices present in the system. Figure \ref{Figure4: Bragg-spectrum} shows the result for both cases. We observed a significant Doppler broadening of the spectrum in the presence of vortices, as a result of the rotation which was investigated in Figure \ref{Figure3: CM Displacements}.

In order to accurately reconstruct the velocity field, one must have a narrow spectral resolution. The initial width of the diffracted stripe of atoms is determined by the linewidth of the Bragg resonance and can be affected by several broadening mechanisms \cite{blak02,stam01bragrev}. Our results in Figure \ref{Figure4: Bragg-spectrum} ({\em solid circles}) show a rms width of 3 kHz for a non-rotating condensate, which is consistent with the residual Doppler shifts due to condensate sloshing (measured to be $\sim$800 $\mu$m/s) as well as the other broadening mechanisms \cite{Footnote1}.

We can calculate the theoretical spectrum by noting that the total number of diffracted atoms at frequency $\delta$ is proportional to the area within the left and right-moving slices for $y=\frac{\lambda (\delta-\delta_o)}{2\Omega}$, as defined in Figure \ref{Figure3: CM Displacements}d. Using a Thomas-Fermi (TF) profile for the density distribution and integrating inside the volume defined by the corresponding TF radii, we find that the spectral lineshape is $\propto \left[\left (\frac{\delta-\delta_o}{A}\right)^2-1\right]^2$, where $A = 2 \Omega R/\lambda$.
The rms width of this distribution is $W_{rms} = 0.38 \times A$, and a fit of the data ({\em squares}) in Figure \ref{Figure4: Bragg-spectrum}, for $|\delta-\delta_o| \leq A$, yields $W_{rms} = 7.5$kHz.  We can compare this with the theoretical estimate for $W$ obtained using $\Omega = 2 \pi \times 16$ Hz, as measured from the data in Figure \ref{Figure3: CM Displacements} and $R = 35 \mu$m.  When convolved with the experimental rms resolution of 3 kHz observed for non-rotating clouds, this yields $W_{rms} = 5.4$ kHz.

\begin{figure}[tbph]
\begin{center}
\vspace{-2mm}
\includegraphics[width = 0.43 \textwidth]{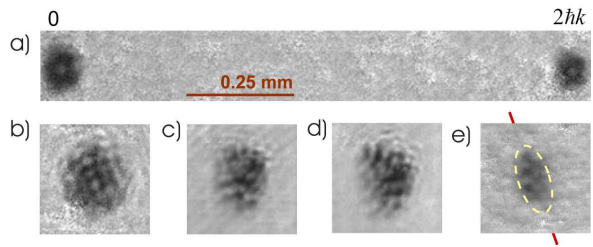}
\end{center}
\vspace{-4mm}
\caption{(Color online) Microscopic structures seen in Bragg scattered clouds. A $T_B$=20 $\mu$s pulse $\delta$=102kHz was used in (a), in a single vortex condensate (at left side of image). The sequence (b)-(e) show images for a rapidly rotating (full lattice) condensate. Experimental parameters are: $T_B$=75$\mu$s, $\delta$=102kHz in (b); $T_B$=60$\mu$s, $\delta$=102kHz in (c); $T_B$=60$\mu$s, $\delta$=122kHz in (d) and $T_B$=50$\mu$s, $\delta$=115kHz in (e). For all images TOF=20 ms. An interesting combination of the quantum (microscopic structures) and classical (rigid body tilting) superfluid flow is shown in (e).}
\vspace{-2mm}
\label{Figure5: Microscopic structures in Bragg}
\end{figure}

A rigidly rotating classical body has a uniform distribution of vorticity, while for a rotating quantum fluid the vorticity is concentrated in individual cores.  The spatial profile of the Bragg scattered atoms should contain evidence of the quantum nature of the rotation.  In Figure 5a, we have observed this quantum signature--the diffracted atom cloud shows a distinct density minimum correlating with that present in the undiffracted atoms.  We obtained this image by holding the rotating BEC for 10 seconds until only 1-2 vortices remained in the gas and by using a short, 20 $\mu$s Bragg pulse with a corresponding broad velocity spread of 2.4 mm/s.  In this limit, the spatial structure of the diffracted atoms is more closely related to the original density profile than to the velocity distribution.  Unlike the TOF images in Figure 1a, these are subject to additional blurring due to motion of the recoiling atoms during the Bragg pulse and therefore the cores have lower contrast.  We could also observe vortex signatures corresponding to an entire lattice, as shown in Figure 5(b-e). A particularly interesting one is 5e where one can see in the same image a combination of quantum and classical fluid behavior. However, the TOF expansion dynamics of the scattered atoms are quite complex due to the mean-field interaction with the undiffracted condensate.  Therefore, while these signatures are intriguing, care must be taken to correlate them with the microscopic structure of the original rotating BEC.


In conclusion, we have used Bragg scattering to directly measure the velocity profile of a rotating BEC.  The technique is complementary to time-of-flight imaging, and might have direct application to the study of non-equilibrium superfluid dynamics \cite{donn91,bare01}, where the tangle of vortices reduces their visibility. We also show that under certain conditions one can clearly observe signatures of the vortices in the diffracted cloud of atoms. Our observations show that while the classical flow is readily understood, further understanding of the time-of-flight dynamics is necessary to provide a quantitative account of the microscopic velocity field.


We thank Brian Kennedy and Leslie Baksmaty for a critical reading of the manuscript and Michael Chapman for loaning a frequency synthesizer used in the experiments.  This work was supported by the U.S. Dept.\ of Energy, the Army Research Office and by Georgia Tech.


\begin{thebibliography}{26}
\bibitem{Footnote1}
In the case of a stationary BEC, the quadrature sum of all the broadening mechanisms (mean-field, the finite pulse duration, finite size effect and light intensity broadening) calculated for our system is $\Delta \nu \cong $ 1 kHz.
\end{thebibliography}

\end{document}